\documentclass[aps,prl,twocolumn,superscriptaddress,amsmath,amssymb,10pt]{revtex4-2}

\usepackage{graphicx}
\usepackage{microtype}
\usepackage{silence}
\usepackage{bm}
\usepackage{hyperref}
\usepackage{xcolor}
\usepackage{pgfplots}
\pgfplotsset{compat=1.18}
\usepgfplotslibrary{groupplots}
\usetikzlibrary{shapes.geometric}

\hypersetup{colorlinks=true,linkcolor=blue!60!black,citecolor=blue!60!black,urlcolor=blue!60!black}

\newcommand{\vect}[1]{\bm{#1}}
\newcommand{\avg}[1]{\langle #1 \rangle}
\newcommand{\abs}[1]{\lvert #1 \rvert}
\newcommand{\sgn}{\operatorname{sgn}}
\newcommand{\order}[1]{\mathcal{O}\!\left(#1\right)}
\newcommand{\dd}{\mathrm{d}}
\newcommand{\Q}{\vect{Q}}
\newcommand{\x}{\vect{x}}
\newcommand{\q}{\vect{q}}
\newcommand{\rr}{\vect{r}}
\newcommand{\vv}{\vect{v}}

\makeatletter
\newcounter{smbibctr}
\newenvironment{smbibliography}[1]{%
  \par\bigskip
  \noindent\textbf{\refname}\par\medskip
  \list{[\arabic{smbibctr}]}%
       {\usecounter{smbibctr}%
        \settowidth\labelwidth{[#1]}%
        \leftmargin\labelwidth
        \advance\leftmargin\labelsep
        \itemsep0pt \parsep0pt}%
  \small\sloppy}%
 {\endlist}
\newcommand{\smbibitem}[1]{\item\phantomsection\label{smbib:#1}}
\newcommand{\smcite}[1]{\textsuperscript{\ref{smbib:#1}}}
\makeatother

\begin{document}


\title{Spectral origin of conformal invariance in active nematic
turbulence}

\author{Rithvik Redrouthu}
\affiliation{Thomas Jefferson High School for Science and Technology,
Alexandria, Virginia 22312, USA}

\date{April 8, 2026}

\begin{abstract}
Zero-vorticity contours in the collective flows of living cells obey
Schramm--Loewner evolution with diffusivity $\kappa = 6$ and thus
fall in the universality class of critical percolation.
This observation is surprising because the underlying vorticity field
has long-range correlations that, according to the
Weinrib--Halperin criterion, should alter the universality class.
Here we propose a spectral explanation for this apparent paradox in
two-dimensional active nematic turbulence.
The universal energy spectrum $E(q) \sim q^{-1}$ implies sign-field
correlations whose decay exponent $a = 3/2$ matches the
Weinrib--Halperin marginal threshold $2/\nu_0 = 3/2$ for
two-dimensional percolation.
At this marginal point the long-range correlations are irrelevant
under renormalization, so the system flows to the uncorrelated
percolation fixed point.
Gaussian surrogate fields with the same spectrum confirm $a = 3/2$
to three significant figures, and left-passage analysis of their
zero-vorticity interfaces yields $\kappa = 5.98 \pm 0.08$,
consistent with $\text{SLE}_6$.
\end{abstract}

\maketitle

Zero-vorticity contours in two-dimensional (2D) turbulent flows can
exhibit conformal invariance, a symmetry more commonly associated
with equilibrium critical phenomena.
Bernard \textit{et al.}~\cite{Bernard2006} first showed that, in the
2D inverse Navier--Stokes cascade, these contours follow
Schramm--Loewner evolution (SLE) with diffusivity $\kappa = 6$,
placing them in the universality class of critical
percolation~\cite{Schramm2000,Smirnov2001,Cardy1992}.
More recently, Andersen \textit{et al.}~\cite{Andersen2025}
found the same $\text{SLE}_6$ statistics in the collective flows of
four living cell systems, ranging from epithelial monolayers to
bacterial suspensions.

At first sight, these results are hard to reconcile with standard
percolation theory because the vorticity field exhibits long-range
spatial correlations.
The Weinrib--Halperin (WH) extended Harris
criterion~\cite{Weinrib1983,Weinrib1984} predicts that such
correlations should generically drive the percolation transition into
a different universality class.
Recent rigorous work has reinforced this expectation, showing that
Gaussian fields with power-law correlations $\sim r^{-a}$ acquire an
effective exponent $\nu_\text{eff} = 2/a$ whenever $a$ lies below the
relevant threshold~\cite{Chalhoub2024}.
The central question is therefore clear: why does uncorrelated
percolation nevertheless survive in these systems?

Eling recently proposed a topological explanation for conformal
invariance in Navier--Stokes turbulence based on Clebsch variables~\cite{Eling2025}.
That explanation, however, relies on the existence of an energy cascade,
whereas active nematics inject and dissipate energy at essentially the
same scale~\cite{Alert2020}.

In this Letter we instead argue for a spectral route to conformal
invariance in 2D active nematic turbulence.
The universal energy spectrum $E(q) \sim q^{-1}$ of this
system~\cite{Alert2020,MartinezPrat2021} implies vorticity sign-field
correlations that decay as $r^{-3/2}$, exactly the WH marginal value
$2/\nu_0 = 3/2$ for 2D percolation ($\nu_0 = 4/3$).
At this point the long-range correlations are marginally irrelevant
under renormalization, so the system flows to the uncorrelated
percolation fixed point, $\text{SLE}_6$.
We test this prediction in Gaussian surrogate fields and then perform
an independent left-passage analysis, both of which support the
$\text{SLE}_6$ interpretation.

\textit{Vorticity correlations.}---In the Stokes regime of 2D active
nematics, the velocity $\vv$ is slaved to the nematic tensor order
parameter $\Q$ through the force balance~\cite{BerisEdwards}
\begin{equation}
  \eta\,\nabla^2 \vv = \nabla p - \alpha\,\nabla\cdot\Q\,,
  \label{eq:stokes}
\end{equation}
where $\eta$ is the viscosity, $\alpha$ is the activity coefficient,
and $p$ enforces incompressibility, $\nabla\cdot\vv = 0$.
The nematic tensor itself evolves according to the Beris--Edwards
dynamics with elastic constant $K$.
Above a critical activity the system enters a turbulent state
populated by $\pm 1/2$ topological defects separated by the active
length scale $\ell_a =
\sqrt{K/\abs{\alpha}}$~\cite{Giomi2015,Doostmohammadi2018,
Hillebrand2025}.

Starting from this force balance, Alert \textit{et al.}~\cite{Alert2020}
showed that the Fourier-space scaling relation
$\eta q^2 \hat{v} \sim \alpha q\,\hat{Q}$, together with a white-noise
spectrum for $\Q$ at wavenumbers $q \lesssim q_a \equiv \ell_a^{-1}$,
leads to the universal energy spectrum
\begin{equation}
  E(q) \sim \frac{\alpha^2}{\eta^2}\,\frac{1}{q}\,,
  \qquad q \lesssim q_a\,,
  \label{eq:spectrum}
\end{equation}
Experiments on microtubule--kinesin active nematics also observe
this scaling~\cite{MartinezPrat2021}.

The corresponding vorticity field,
$\omega = (\nabla\times\vv)\cdot\hat{\vect{z}}$,
has enstrophy spectrum $\Omega(q) = q^2 E(q) \sim q$, so the spectrum
grows linearly up to the active wavenumber $q_a$.
To translate this statement into real space, we take the inverse
Hankel transform of $\Omega(q)$.
Using the identity $\int_0^w u\,J_0(u)\,\dd u =
w\,J_1(w)$~\cite{Watson1944}, we obtain
\begin{equation}
  C_\omega(r) = A\,\frac{q_a}{r}\,J_1(q_a r)\,,
  \label{eq:Comega}
\end{equation}
with normalized correlation $\rho(r) = 2J_1(q_a r)/(q_a r)$.
For $r \gg \ell_a$, the asymptotic form
$J_1(x) \sim \sqrt{2/(\pi x)}\cos(x-3\pi/4)$ shows that the
correlation oscillates while its envelope decays as $r^{-3/2}$.

\textit{Sign-field exponent and spectral marginality.}---The
zero-vorticity contours are the domain walls of the binary sign
field $\sigma(\x) = \sgn[\omega(\x)]$, which divides the plane into
regions of positive and negative vorticity.
Because the turbulent state is statistically symmetric under
$\omega \to -\omega$~\cite{Alert2020}, we have $\avg{\sigma} = 0$,
so the system sits exactly at the site-percolation
threshold~\cite{Isichenko1992,Bogomolny2002}.
The universality class of the contour ensemble is therefore set by the
correlation structure of $\sigma$.

For a Gaussian random field, that correlation is given exactly by the
arcsine law~\cite{VanVleck1943}:
\begin{equation}
  C_\sigma(r) = \frac{2}{\pi}\arcsin\!\left(\rho(r)\right)
  \approx \frac{2}{\pi}\,\rho(r) \sim r^{-3/2}
  \quad (r \gg \ell_a)\,,
  \label{eq:arcsine}
\end{equation}
which immediately yields the sign-field exponent $a = 3/2$.
Active-nematic vorticity is not itself Gaussian, since it is produced
by a dilute gas of topological defects~\cite{Giomi2015,Doostmohammadi2018},
so higher connected cumulants can in principle modify the arcsine
prediction.
For a dilute defect gas, however, the leading correction is governed
by the connected fourth cumulant $\kappa_4(r)$, which decays as
$r^{-\beta}$ with $\beta \geq 2$~\cite{SM}.
Because this decay is faster than the Gaussian contribution
$r^{-3/2}$, the long-distance exponent remains controlled by the
two-point function.
The value $a = 3/2$ is therefore robust to the non-Gaussian structure
of the turbulence, provided the universal spectrum
$E(q) \sim q^{-1}$ is maintained~\cite{Alert2020}.

We can now compare this exponent with the WH extended Harris
criterion~\cite{Weinrib1983,Weinrib1984}.
That criterion states that site correlations decaying as $r^{-a}$ are
relevant for $a < 2/\nu_0$ and irrelevant for $a > 2/\nu_0$, where
$\nu_0$ is the correlation-length exponent of uncorrelated
percolation.
In two dimensions, $\nu_0 = 4/3$
exactly~\cite{Smirnov2001,Stauffer1994}, so
\begin{equation}
  \frac{2}{\nu_0} = \frac{3}{2} = a\,.
  \label{eq:marginal}
\end{equation}
Thus the sign-field exponent of active nematic turbulence lands
precisely at the WH marginal threshold.
The exponent $a = 3/2$ is fixed by Stokes-regime
kinematics~\cite{Alert2020} and $\nu_0 = 4/3$ by the conformal
field theory of 2D percolation~\cite{Cardy1992}.
Neither quantity depends on microscopic material parameters.

Although the vorticity correlation in Eq.~\eqref{eq:Comega} oscillates, the WH criterion is controlled by the long-distance scaling of the disorder variance, which depends only on the envelope exponent $a$ and is insensitive to the oscillatory structure~\cite{Weinrib1983}.

At marginality, the WH $\beta$-function for the long-range disorder
coupling $g$ reduces to~\cite{Weinrib1984,Prudnikov2000}
$\beta_g = -c_1 g^2 + \order{g^3}$,
so the running coupling
$g(\ell) = g_0/(1 + c_1 g_0 \ln(\ell/\ell_0))$
flows logarithmically to zero.
The theory therefore approaches the uncorrelated percolation fixed
point, namely $\text{SLE}_6$.
It is also useful to clarify the spectral origin of this exponent.
The $r^{-3/2}$ envelope in Eq.~\eqref{eq:Comega} is tied to the
spectral discontinuity at $q_a$.
If the cutoff is smoothed, the decay becomes even faster; for example,
an exponential rolloff yields $a = 3$~\cite{SM}.
Since $a > 3/2$ lies in the WH \textit{irrelevant} regime, any
physical softening of the cutoff only drives the system further into
the $\text{SLE}_6$ class rather than away from it.

\begin{figure*}[t]
  \centering
  \begin{minipage}[b]{0.54\textwidth}
    \begin{tikzpicture}
    \begin{axis}[
      width=\textwidth, height=6.2cm,
      xmode=log, ymode=log,
      xlabel={$r$ [pixels]}, ylabel={$|C_\sigma(r)|$},
      xmin=0.8, xmax=500, ymin=3e-5, ymax=2,
      tick align=inside,
      label style={font=\normalsize},
      tick label style={font=\small},
      legend style={font=\small, draw=none, fill=none,
                    at={(0.03,0.03)}, anchor=south west,
                    cells={anchor=west}, row sep=0pt},
      major tick length=4pt, minor tick length=2.5pt,
    ]
    \addplot[black, no markers, line width=0.5pt]
      table[x=r, y=csigma]{fig1a_data.dat};
    \addlegendentry{measured}
    \addplot[red!80!black, dashed, no markers, line width=0.7pt]
      table[x=r, y=canalytic]{fig1a_data.dat};
    \addlegendentry{arcsine law}
    \addplot[green!50!black, no markers, line width=0.7pt]
      table[x=r, y=env]{fig1a_envelope.dat};
    \addlegendentry{$r^{-3/2}$ envelope}
    \node[font=\bfseries\Large] at (rel axis cs:0.07,0.94) {a};
    \end{axis}
    \end{tikzpicture}
  \end{minipage}%
  \hfill
  \begin{minipage}[b]{0.44\textwidth}
    \begin{tikzpicture}
    \begin{axis}[
      width=\textwidth, height=6.2cm,
      xmode=log,
      xlabel={$L/\ell_a$}, ylabel={sign-field exponent $a$},
      xmin=25, xmax=500, ymin=1.30, ymax=1.55,
      xtick={30,100,300},
      xticklabels={30,100,300},
      tick align=inside,
      label style={font=\normalsize},
      tick label style={font=\small},
      major tick length=4pt, minor tick length=2.5pt,
    ]
    \draw[red!80!black, dashed, line width=0.7pt]
      (axis cs:25,1.5) -- (axis cs:500,1.5);
    \node[red!80!black, font=\small, anchor=south west]
      at (axis cs:180,1.502) {$a = 3/2$};
    \addplot[only marks, mark=*, mark size=2.5pt, black]
      table[x=Lla, y=a]{fig1b_convergence.dat};
    \node[font=\bfseries\Large] at (rel axis cs:0.07,0.94) {b};
    \end{axis}
    \end{tikzpicture}
  \end{minipage}
  \caption{%
  \textbf{Sign-field exponent in Gaussian surrogates.}
  (a)~Absolute value of the sign-field correlation $|C_\sigma(r)|$
  at $N = 3072$ ($L/\ell_a = 384$), shown together with the arcsine
  law $(2/\pi)\arcsin(\rho(r))$ (dashed red) and a power-law envelope
  with $a = 3/2$ (green). The agreement extends across three decades
  in $r$.
  (b)~Convergence of the measured sign-field exponent $a$ toward the
  predicted value $3/2$ (dashed line) as the system size increases,
  shown for $\gamma = 1$ with 160 realizations per size.}
  \label{fig:tier2}
\end{figure*}
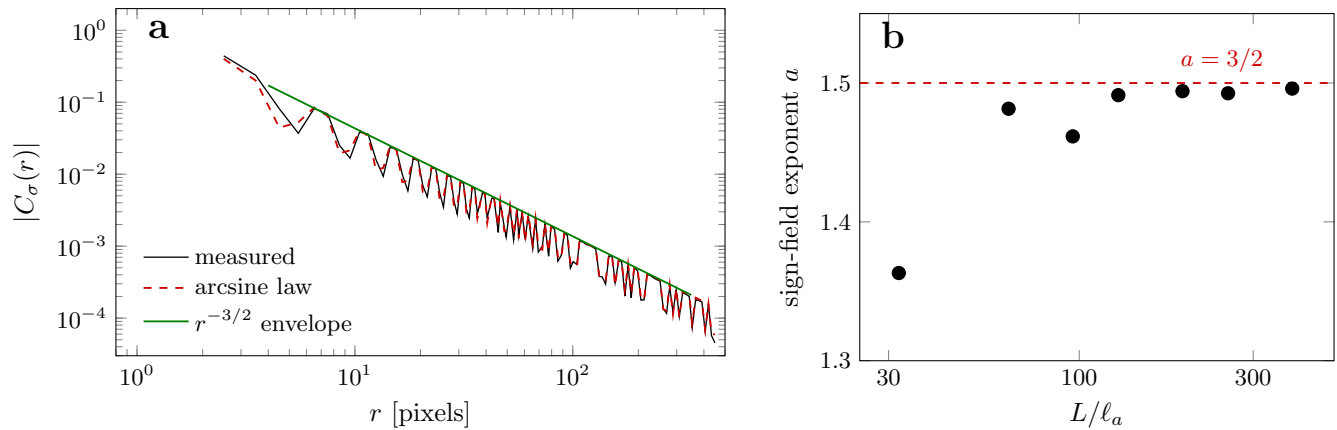

\textit{Surrogate validation.}---We test the central prediction
$a = 3/2$ using Gaussian random fields with prescribed energy
spectrum $E(q) \sim q^{-\gamma}$ and a sharp cutoff at $q_a$.
By construction, this surrogate isolates the spectral mechanism from
defect-specific physics, so it provides a clean test of the argument
even though a full active-nematic verification remains an important
future goal.
We work on periodic grids with $N = 256$--$3072$
($L/\ell_a = 32$--$384$), fix $\ell_a = 8$ pixels, and average over
160 independent realizations at each size~\cite{SM}.

For the active-nematic case $\gamma = 1$, the surrogate reproduces
the predicted marginal exponent.
At the largest size, $N = 3072$, we find $a = 1.4960$, in excellent
agreement with the theoretical value $a = 3/2$
(Fig.~\ref{fig:tier2}).
Moreover, the measured sign-field correlation follows the arcsine law
over three decades in separation and four decades in amplitude, while
the $r^{-3/2}$ envelope is confirmed to three significant figures.

\textit{Spectral threshold.}---The same analysis extends naturally
beyond the active-nematic case $E(q) \sim q^{-1}$.
For a general enstrophy spectrum $\Omega(q) \sim q^\mu$, the
sign-field exponent is controlled by whether the
Weber--Schafheitlin integral $\int_0^\infty u^\mu J_0(u)\,\dd u$
converges~\cite{Watson1944,SM}.
If $\mu < 1/2$, the integral converges and one finds
$a = 1 + \mu$.
If $\mu > 1/2$, the integral diverges and the Hankel transform is
instead governed by oscillations whose envelope decays universally as
$r^{-3/2}$, so $a = 3/2$ becomes independent of $\mu$~\cite{SM}.
The crossover therefore occurs at $\mu_c = 1/2$, corresponding to
$E(q) \sim q^{-3/2}$.

\begin{figure}[t]
  \centering
  \begin{minipage}[b]{\columnwidth}
    \begin{tikzpicture}
    \begin{axis}[
      width=\columnwidth, height=5cm,
      xlabel={spectral exponent $\gamma$},
      ylabel={sign-field exponent $a$},
      xmin=0.3, xmax=2.15, ymin=1.1, ymax=1.58,
      tick align=inside,
      label style={font=\normalsize},
      tick label style={font=\small},
      major tick length=4pt,
      xtick={0.5,1.0,1.5,2.0},
    ]
    \draw[red!80!black, dashed, line width=0.7pt]
      (axis cs:0.3,1.5) -- (axis cs:2.15,1.5);
    \node[red!80!black, font=\small, anchor=south east]
      at (axis cs:2.1,1.502) {$a = 3/2$};
    \draw[gray, dotted, line width=0.5pt]
      (axis cs:1.5,1.1) -- (axis cs:1.5,1.58);
    \node[gray, font=\small, anchor=south west]
      at (axis cs:1.52,1.11) {$\gamma_c$};
    \addplot[only marks, mark=*, mark size=2.5pt, black]
      table[x=gamma, y=a]{fig2a_threshold.dat};
    \node[font=\bfseries\Large] at (rel axis cs:0.06,0.93) {a};
    \end{axis}
    \end{tikzpicture}
  \end{minipage}\\[6pt]
  \begin{minipage}[b]{\columnwidth}
    \begin{tikzpicture}
    \begin{axis}[
      width=\columnwidth, height=5cm,
      ylabel={estimated $\kappa$},
      xmin=0.4, xmax=2.6, ymin=5.5, ymax=6.4,
      xtick={0.85, 2.0},
      xticklabels={$N = 4096$, $N = 6144$},
      ytick={5.5,5.6,...,6.4},
      minor y tick num=1,
      tick align=inside,
      label style={font=\normalsize},
      tick label style={font=\small},
      legend style={font=\small, at={(0.03,0.03)},
                    anchor=south west, draw=none, fill=none},
      major tick length=4pt,
    ]
    \draw[gray, dashed, line width=0.5pt]
      (axis cs:0.4,6) -- (axis cs:2.6,6);
    \node[gray, font=\small, anchor=south west]
      at (axis cs:2.15,6.01) {$\kappa = 6$};
    \addplot[only marks, mark=*, mark size=2.5pt,
      green!50!black, fill=green!50!black,
      error bars/.cd, y dir=both, y explicit,
      error bar style={line width=0.6pt, green!50!black},
      error mark options={rotate=90, mark size=2.5pt,
        line width=0.5pt, green!50!black},
    ] coordinates { (0.75, 6.034) +- (0, 0.114) (1.88, 5.922) +- (0, 0.127) };
    \addlegendentry{Gaussian field}
    \addplot[only marks, mark=square*, mark size=2pt,
      orange!80!black, fill=orange!80!black,
      error bars/.cd, y dir=both, y explicit,
      error bar style={line width=0.6pt, orange!80!black},
      error mark options={rotate=90, mark size=2.5pt,
        line width=0.5pt, orange!80!black},
    ] coordinates { (0.95, 5.835) +- (0, 0.111) (2.08, 5.916) +- (0, 0.202) };
    \addlegendentry{percolation control}
    \node[font=\small, draw=gray!50, rounded corners=1.5pt,
      fill=white, inner sep=3pt, line width=0.3pt,
      anchor=north east]
      at (axis cs:2.55,6.39)
      {pooled: $\kappa = 5.984 \pm 0.085$};
    \node[font=\bfseries\Large] at (rel axis cs:0.06,0.93) {b};
    \end{axis}
    \end{tikzpicture}
  \end{minipage}
  \caption{%
  \textbf{Spectral threshold and left-passage check.}
  (a)~Sign-field exponent $a$ as a function of spectral slope
  $\gamma$ at $N = 3072$.
  For $\gamma < \gamma_c = 3/2$ (dotted line), the measurements cluster
  near the marginal value $3/2$ (dashed line), whereas above the
  threshold the exponent trends downward.
  (b)~SLE diffusivity $\kappa$ extracted from boundary-conditioned
  zero-vorticity interfaces in Gaussian surrogates (circles) and in a
  triangular-lattice percolation control (squares).
  Pooling the two largest sizes gives
  $\kappa = 5.98 \pm 0.08$, consistent with $\text{SLE}_6$.}
  \label{fig:tier1}
\end{figure}
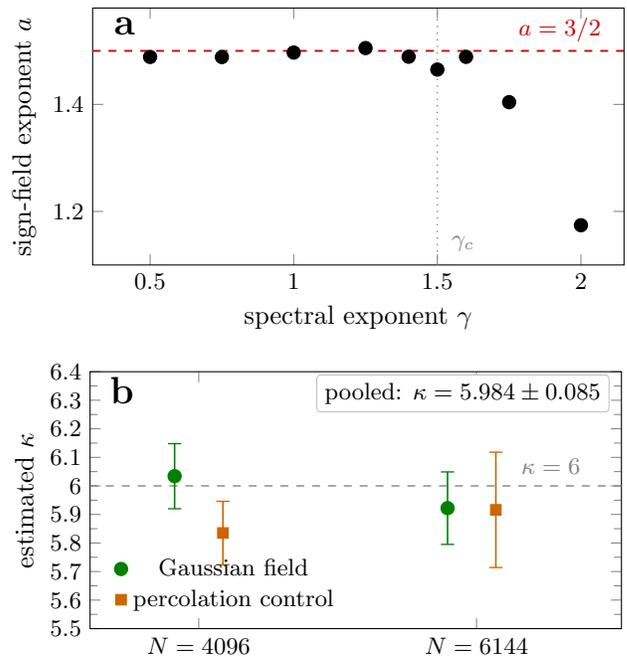

The numerical results at $N = 3072$ follow this pattern closely.
Below the threshold, the sign-field exponent stays near $3/2$
($a = 1.489$ at $\gamma = 0.50$, $a = 1.497$ at $\gamma = 1.00$,
$a = 1.505$ at $\gamma = 1.25$), whereas well above the threshold it
drops substantially
($a = 1.404$ at $\gamma = 1.75$, $a = 1.174$ at
$\gamma = 2.00$) (Fig.~\ref{fig:tier1}a)~\cite{SM}.
Finite-size effects broaden the crossover region, but the data remain
consistent with $\gamma_c = 3/2$.

This threshold also clarifies how active nematics differ from the
classical inverse cascade.
For the Kolmogorov spectrum $E(q) \sim q^{-5/3}$,
$\mu = 1/3 < 1/2$, so $a_\text{IC} = 4/3 < 3/2$ and the long-range
correlations are \textit{relevant} rather than marginal.
The spectral-marginality mechanism therefore does not apply in that
regime, and the $\text{SLE}_6$ behavior observed by Bernard
\textit{et al.}~\cite{Bernard2006} likely requires a different
explanation~\cite{Falkovich2009,Eling2025}.
By contrast, active nematics lie on the $\mu > 1/2$ side of the
threshold, where marginality is robust.
This observation leads to a concrete experimental prediction:
systems whose spectrum is steepened beyond $E(q) \sim q^{-3/2}$---for
example by substrate friction~\cite{MartinezPrat2021}---should leave
the marginal regime and lose conformal invariance.

\textit{Left-passage check.}---As an independent geometric test, we
analyze boundary-conditioned zero-vorticity interfaces using
Schramm's left-passage probability~\cite{Schramm2001,SM}.
We extract exploration paths from the sign field in a disk geometry
at $N = 4096$ and $N = 6144$, and then fit the resulting statistics to
the left-passage formula to estimate $\kappa$.
We obtain $\kappa = 6.03 \pm 0.11$ from 960 interfaces at
$N = 4096$ and $\kappa = 5.92 \pm 0.13$ from 640 interfaces at
$N = 6144$.
Combining the two datasets gives $\kappa = 5.98 \pm 0.08$, which is
fully consistent with $\text{SLE}_6$ (Fig.~\ref{fig:tier1}b).
As a check on the analysis pipeline, we apply the same procedure to a
positive control: critical site percolation on a triangular lattice
($p_c = 1/2$), rasterized in the same disk geometry~\cite{SM}.
For that control we find $\kappa = 5.84 \pm 0.11$ at $N = 4096$
and $\kappa = 5.92 \pm 0.20$ at $N = 6144$, again consistent with
$\kappa = 6$ at the present level of precision.

\textit{Discussion.}---Taken together, these results suggest a direct
link between the universal energy spectrum of active nematic
turbulence and the conformal invariance of its zero-vorticity
contours.
The logic proceeds in three steps: the Stokes-regime spectrum
$E(q) \sim q^{-1}$~\cite{Alert2020} fixes the sign-field exponent at
$a = 3/2$ through the arcsine law; that value coincides with the WH
marginality threshold for 2D percolation; and the resulting
logarithmic RG flow drives the system toward the uncorrelated fixed
point, $\text{SLE}_6$~\cite{Smirnov2001,Weinrib1984,VanVleck1943}.
The Gaussian surrogates validate the first step to three significant
figures, while the left-passage analysis independently supports the
final conclusion with $\kappa = 5.98 \pm 0.08$.

We emphasize, however, that the surrogate fields validate the
spectral mechanism rather than the full active-nematic dynamics.
As a first step beyond the Gaussian approximation, we reanalyzed the
experimental PIV data of Andersen \textit{et al.}~\cite{Andersen2025}.
For MDCK monolayers we find $a = 1.53 \pm 0.20$, while MCF-7 breast
cancer cells yield $a = 1.44 \pm 0.02$~\cite{SM}.
The averaged energy spectra in both datasets also display an
intermediate-wavenumber band with slope near $-1$.
Together with the published observation $\kappa \approx 6$, these
measurements provide preliminary, though not yet definitive, support for the
spectral-marginality mechanism in non-Gaussian experimental flows.
A definitive test will require larger fields of view and
direct Beris--Edwards simulations.

The threshold $\gamma_c = 3/2$ also leads to a sharp falsifiable
prediction.
If substrate friction steepens the spectrum in
microtubule--kinesin nematics~\cite{MartinezPrat2021}, then the
system should move out of the marginal regime and show a measurable
departure from $\kappa = 6$.
More broadly, the framework distinguishes 2D turbulent systems that
are perturbatively accessible through spectral marginality---namely
those with $\mu > 1/2$, including active nematics---from those that
likely require non-perturbative explanations~\cite{Falkovich2009,Eling2025}.
Whether related spectral coincidences underlie SLE statistics in other
driven systems, such as weakly compressible
turbulence~\cite{Puggioni2020} or gravity-wave
turbulence~\cite{Bernard2007,Noseda2024}, remains an open question,
but the present framework makes that question quantitatively precise.


\setcounter{equation}{0}
\setcounter{figure}{0}
\setcounter{table}{0}
\setcounter{section}{0}
\renewcommand{\theequation}{S\arabic{equation}}
\renewcommand{\thesection}{S\Roman{section}}
\renewcommand{\thefigure}{S\arabic{figure}}
\renewcommand{\thetable}{S\arabic{table}}
\renewcommand{\theHequation}{S\arabic{equation}}
\renewcommand{\theHsection}{S\Roman{section}}
\renewcommand{\theHfigure}{S\arabic{figure}}
\renewcommand{\theHtable}{S\arabic{table}}

\clearpage
\onecolumngrid
\vspace*{-\baselineskip}
\begin{center}
{\Large Supplemental Material for ``Spectral origin of conformal invariance in active nematic turbulence''}

\bigskip
Rithvik Redrouthu\textsuperscript{1}

\smallskip
{\small \textsuperscript{1}\textit{Thomas Jefferson High School for Science and Technology,
Alexandria, Virginia 22312, USA}}

\smallskip
(Dated: \today)
\end{center}
\bigskip
\twocolumngrid

\section{Hankel transform asymptotics}
\label{sec:hankel}

\subsection{Setup}

For a 2D isotropic velocity field with energy spectrum $E(q)$, the
enstrophy spectrum is $\Omega(q) = q^2 E(q)$.
The vorticity correlation function is
\begin{equation}
  C_\omega(r) = A \int_0^{q_a} q^\mu\,J_0(qr)\,\dd q\,,
  \label{eq:Cgeneral}
\end{equation}
where $\Omega(q) \propto q^\mu$ for $q \leq q_a$ and $A$ absorbs
all prefactors.
For active nematics ($E \sim q^{-1}$), $\mu = 1$; for the inverse
cascade ($E \sim q^{-5/3}$), $\mu = 1/3$.

\subsection{\texorpdfstring{Exact result for $\mu = 1$}{Exact result for mu = 1}}

Substituting $u = qr$ and using
$\frac{\dd}{\dd u}[uJ_1(u)] = uJ_0(u)$, hence
$\int_0^w uJ_0(u)\,\dd u = wJ_1(w)$~\smcite{Watson1944}:
\begin{equation}
  C_\omega(r) = A\,\frac{q_a}{r}\,J_1(q_a r)\,.
  \label{eq:Cactive}
\end{equation}
Using $\lim_{u\to 0}J_1(u)/u = 1/2$, the zero-lag value is
$C_\omega(0) = Aq_a^2/2$, and the normalized correlation is
$\rho(r) = 2J_1(q_a r)/(q_a r)$.
For $q_ar \gg 1$, $J_1(x) \sim \sqrt{2/(\pi x)}\cos(x - 3\pi/4)$
gives
\begin{equation}
  \rho(r) \sim \sqrt{\frac{8}{\pi}}\,(q_ar)^{-3/2}\cos(\cdots)\,,
  \label{eq:rho_asymp}
\end{equation}
confirming $a = 3/2$.

\subsection{\texorpdfstring{Convergent regime: $\mu < 1/2$}{Convergent regime: mu < 1/2}}

Writing $C_\omega(r) = A\,r^{-(1+\mu)}I_\mu(q_ar)$ with
$I_\mu(w) \equiv \int_0^w u^\mu J_0(u)\,\dd u$, the large-$r$
behavior is controlled by $I_\mu(w \to \infty)$.
For $-1 < \mu < 1/2$, the Weber--Schafheitlin
formula~\smcite{Watson1944} gives
\begin{equation}
  L_\mu \equiv \int_0^\infty u^\mu J_0(u)\,\dd u
  = \frac{2^\mu\,\Gamma\!\bigl(\frac{1+\mu}{2}\bigr)}
  {\Gamma\!\bigl(\frac{1-\mu}{2}\bigr)}\,,
  \label{eq:WS}
\end{equation}
which converges (conditionally for $\mu > -1/2$, absolutely for
$\mu < -1/2$).
Therefore $C_\omega(r) \to AL_\mu\,r^{-(1+\mu)}$ and
\begin{equation}
  a = 1 + \mu \qquad (\mu < \tfrac{1}{2})\,.
  \label{eq:a_conv}
\end{equation}
For the inverse cascade ($\mu = 1/3$):
$a_\text{IC} = 4/3 < 3/2 = 2/\nu_0$.

\subsection{\texorpdfstring{Divergent regime: $\mu > 1/2$}{Divergent regime: mu > 1/2}}

When $\mu > 1/2$, $\int_0^\infty u^\mu J_0(u)\,\dd u$ diverges.
Using the identity
\begin{equation}
  \int_0^w u^\mu J_0(u)\,\dd u = w^\mu J_1(w)
  - (\mu-1)\int_0^w u^{\mu-1}J_1(u)\,\dd u\,,
  \label{eq:IBP}
\end{equation}
(from $\frac{\dd}{\dd u}[u^\mu J_1(u)] = u^\mu J_0(u) +
(\mu-1)u^{\mu-1}J_1(u)$), the first term has envelope
$\sim w^{\mu-1/2}$ while the remainder generates terms
$\sim w^{\mu-3/2}, \ldots$.
For $\mu > 1/2$ the first term dominates, giving
\begin{equation}
  a = \frac{3}{2} \qquad (\mu > \tfrac{1}{2})\,,
  \label{eq:a_div}
\end{equation}
independent of $\mu$.
This is verified at $\mu = 1$: Eq.~\eqref{eq:Cactive} has envelope
$\sim q_a^{1/2}r^{-3/2}$, consistent with $q_a^{\mu-1/2} =
q_a^{1/2}$.

\subsection{Oscillatory correlations and the WH criterion}

The vorticity correlation~\eqref{eq:Cactive} oscillates on the scale
$\ell_a$.
The WH criterion was formulated for monotonic correlations.
At RG scale $\ell \gg \ell_a$, the coarse-grained sign-field
correlation averages over oscillations of period $\ell_a \ll \ell$,
and only the envelope $\sim r^{-3/2}$ survives.
The effective exponent entering the WH criterion is therefore
$a = 3/2$.
This reasoning applies whenever $\xi \gg \ell_a$, which holds near
the percolation threshold where $\xi \to \infty$.

\subsection{Effect of smooth spectral cutoffs}

The $r^{-3/2}$ envelope in Eq.~\eqref{eq:Cactive} is an endpoint
effect: it arises from the spectral discontinuity at $q = q_a$.
Smoothing the cutoff damps these oscillations and produces faster
decay.
For example, with an exponential rolloff
$\Omega(q) = q\,e^{-q/q_a}$, the correlation evaluates
exactly~\smcite{Watson1944}:
\begin{equation}
  \begin{aligned}
    C_\omega(r) &= A\int_0^\infty q\,e^{-q/q_a}J_0(qr)\,\dd q \\
    &= \frac{A\,q_a^{-1}}{(r^2 + q_a^{-2})^{3/2}} \\
    &\sim r^{-3}\,.
  \end{aligned}
\end{equation}
giving sign-field exponent $a = 3$.
Since $a = 3 > 3/2 = 2/\nu_0$, the correlations are
\textit{irrelevant} under the WH criterion, and the system is
squarely in the uncorrelated percolation class ($\text{SLE}_6$)
without requiring the marginal RG flow.
The hard cutoff is therefore the \textit{worst case}: it produces
the slowest-decaying correlations ($a = 3/2$, marginal), whereas
any smooth rolloff gives faster decay ($a > 3/2$, irrelevant),
pushing the system further into the $\text{SLE}_6$ regime.

\section{The arcsine law}
\label{sec:arcsine}

For a bivariate Gaussian $(X,Y)$ with $\avg{X} = \avg{Y} = 0$,
$\avg{X^2} = \avg{Y^2} = 1$, and $\avg{XY} = \rho$, the
orthant probability is~\smcite{VanVleck1943}
\begin{equation}
  P(X > 0,\, Y > 0) = \tfrac{1}{4} +
  \tfrac{1}{2\pi}\arcsin(\rho)\,.
  \label{eq:orthant}
\end{equation}
Since $C_\sigma = 4P(\omega_1>0,\omega_2>0) - 1$ (by the $\omega
\to -\omega$ symmetry):
\begin{equation}
  C_\sigma(r) = \frac{2}{\pi}\arcsin\!\bigl(\rho(r)\bigr)\,.
  \label{eq:arcsine_derived}
\end{equation}
For $\abs{\rho} \ll 1$:
$\arcsin(\rho) = \rho + \rho^3/6 + 3\rho^5/40 + \cdots$, so
$C_\sigma \approx (2/\pi)\rho$.
The cubic correction $\sim \rho^3 \sim r^{-9/2}$ is negligible
compared to $\rho \sim r^{-3/2}$.

\section{Non-Gaussian corrections from the defect gas}
\label{sec:defect}

The active nematic vorticity can be modeled schematically as a dilute
sum over defects at random positions $\x_i$ with random orientations
$\phi_i$,
$\omega(\x) = \sum_i A_d f_i\,K(\x-\x_i)$, where $A_d \sim \alpha/\eta$,
$f_i$ is a bounded angular factor, and the core-regularized far-field
kernel obeys $K(r) \sim 1/\max(r,\ell_a)$~\smcite{Giomi2015}.
The two-point correlation and spectrum $E(q) \sim q^{-1}$ are taken
from the Stokes-regime analysis of Alert \textit{et al.}~\smcite{Alert2020};
the defect-gas model is used here only to estimate non-Gaussian
corrections.

For independent dilute defects, the connected bivariate fourth cumulant
$\kappa_4(r) = \avg{\omega(\x)^2\omega(\x+\rr)^2}_c$ is dominated by
single-defect terms,
\begin{equation}
  \kappa_4(r) \sim \rho_d A_d^4 \int \dd^2 y\,K(y)^2 K(y-r)^2\,.
\end{equation}
For $r \gg \ell_a$, regions within $\order{\ell_a}$ of either
observation point give one factor $K^2 \sim \ell_a^{-2}$ and the other
$K^2 \sim r^{-2}$, hence a contribution $\lesssim r^{-2}$; integrating
over intermediate scales adds at most a logarithm. Thus
\begin{equation}
  \kappa_4(r) \lesssim r^{-2}\ln(r/\ell_a)\,.
\end{equation}
Midpoint contributions decay faster, $\sim r^{-4}$.
The non-Gaussian correction therefore falls at least as fast as
$r^{-2}$ (up to logarithms), so
$\Delta_\text{NG}(r) \sim \kappa_4(r)/C_\omega(0)^2$ remains
subdominant to the Gaussian term $\sim r^{-3/2}$.

\section{Numerical methods}
\label{sec:numerics}

\subsection{Gaussian surrogate field generation}

We generate 2D isotropic Gaussian random fields $\omega(\x)$ on
$N \times N$ periodic grids with prescribed per-mode power spectral
density $S_\omega(q) \equiv \avg{\abs{\hat{\omega}(q)}^2}
\propto q^{1-\gamma}$ for $q \leq q_a = 2\pi/\ell_a$
and $S_\omega = 0$ for $q > q_a$, where $\ell_a = 8$ pixels.
Since the 1D enstrophy spectrum is $\Omega(q) \propto q\,S_\omega(q)
\propto q^{2-\gamma}$, this corresponds to energy spectrum
$E(q) = \Omega(q)/q^2 \sim q^{-\gamma}$.
For $\gamma = 1$ (the active nematic case), $S_\omega \sim q^0$
(white) and $\Omega \sim q$.

The field is constructed by: (1) generating complex Gaussian white
noise $\xi(\q)$ in Fourier space; (2) applying the spectral filter
$\hat{\omega}(\q) = h(|\q|)\,\xi(\q)$ with $h(q) = q^{(1-\gamma)/2}$
for $0 < q \leq q_a$ and $h = 0$ otherwise (and $h(0) = 0$ to
enforce $\avg{\omega} = 0$), so that
$\avg{\abs{\hat{\omega}}^2} = h^2 = q^{1-\gamma}$ as required;
(3) enforcing Hermitian symmetry; (4) inverse-FFT.
Grid sizes are $N = 256$--$6144$, with
$\ell_a = 8$ pixels giving $L/\ell_a = 32$--$768$.

\subsection{Sign-field correlation}

The sign field $\sigma(\x) = \sgn[\omega(\x)]$ is computed
pointwise.
The two-point correlation $C_\sigma(r) = \avg{\sigma(\x+\rr)
\sigma(\x)}$ is computed by FFT-based autocorrelation (Wiener--Khinchin
theorem) and azimuthal averaging.
The exponent $a$ is extracted by fitting $|C_\sigma(r)|$ to
$Ar^{-a}$ over the range $r/\ell_a \in [3, 50]$, where the envelope
is well-developed and finite-size effects are negligible.
For $\gamma = 1$ the surrogate exponent converges monotonically
with system size: $a = 1.35$ ($N = 256$), $1.43$ ($N = 512$),
$1.47$ ($N = 1024$), $1.49$ ($N = 2048$), $1.496$ ($N = 3072$),
reaching the theoretical value $3/2$ to three significant
figures at the largest size (160 realizations per size).

\subsection{SLE diffusivity from left-passage probability}

For each realization, we extract the percolation exploration path on
the sign field within a disk of radius $R = N/3$ centered on the
grid, with boundary conditions set by the continuous field on the
disk boundary.
The exploration path traces the interface between $+$ and $-$
domains from one boundary point into the disk interior.

We measure the left-passage probability $P_\text{left}(z)$ for test
points $z = x + iy$ in the upper half-plane. The statistic is sampled
at 13 values of the aspect ratio $x/y$ uniformly spaced in
$[-1.5, 1.5]$, using all realizations.
The SLE parameter $\kappa$ is extracted by weighted least-squares
fitting to Schramm's formula~\smcite{Schramm2001}:
\begin{equation}
  P_\text{left}(z) = \frac{1}{2} + \frac{\Gamma(4/\kappa)}
  {\sqrt{\pi}\,\Gamma\bigl(\frac{8-\kappa}{2\kappa}\bigr)}\,
  {}_2F_1\!\Bigl(\frac{1}{2}, \frac{4}{\kappa}; \frac{3}{2};
  -\frac{x^2}{y^2}\Bigr)\frac{x}{y}\,.
\end{equation}
Bootstrap resampling (1000 resamples) over the realization ensemble
provides the standard error on $\kappa$.

\subsection{Pipeline validation controls}

We validate the $\kappa$-extraction pipeline with a same-geometry
positive control, namely critical site percolation on the triangular
lattice ($p_c = 1/2$), rasterized at $N = 4096$ and $N = 6144$ on
the same disk geometry used for the Gaussian surrogates.
Exploration paths are extracted and analyzed with the identical
pipeline.
Since critical site percolation on the triangular lattice
is rigorously described by $\text{SLE}_6$~\smcite{Smirnov2001}, this
provides an end-to-end test of the measurement stack (contour
extraction, disk geometry, and left-passage fit) in a system where
$\kappa = 6$ is exact.

Results: $\kappa = 5.84 \pm 0.11$ at $N = 4096$ (960 interfaces)
and $\kappa = 5.92 \pm 0.20$ at $N = 6144$ (320 interfaces), both
consistent with $\kappa = 6$ at the current precision.

\subsection{Spectral threshold data}

Table~\ref{tab:threshold} reports the measured sign-field exponent
$a$ as a function of the spectral slope $\gamma$ at $N = 3072$ (160
realizations per point).

\begin{table}[t]
  \caption{\label{tab:threshold}Sign-field exponent $a(\gamma)$ at
  $N = 3072$. For $\gamma < 3/2$ ($\mu > 1/2$), the measured $a$
  clusters near $3/2$ (marginal regime). For $\gamma > 3/2$
  ($\mu < 1/2$), $a$ drops below $3/2$ (relevant regime), consistent
  with the predicted threshold $\gamma_c = 3/2$.}
  \begin{ruledtabular}
  \begin{tabular}{c c c}
    $\gamma$ & $\mu = 2-\gamma$ & measured $a$ \\
    0.50 & 1.50 & 1.489 \\
    0.75 & 1.25 & 1.488 \\
    1.00 & 1.00 & 1.497 \\
    1.25 & 0.75 & 1.505 \\
    1.40 & 0.60 & 1.489 \\
    1.50 & 0.50 & 1.465 \\
    1.60 & 0.40 & 1.489 \\
    1.75 & 0.25 & 1.404 \\
    2.00 & 0.00 & 1.174 \\
  \end{tabular}
  \end{ruledtabular}
\end{table}

\section{Experimental reanalysis of cell-flow PIV data}
\label{sec:experimental}

As a test of the spectral-marginality mechanism beyond the
Gaussian surrogate approximation, we reanalyzed the publicly
available PIV data from Andersen \textit{et
al.}~\smcite{Andersen2025}.
For each velocity frame we computed the vorticity
$\omega = \partial_x v_y - \partial_y v_x$, constructed the sign
field $\sigma(\x) = \sgn\,\omega(\x)$, and measured the
isotropic sign-field correlation $C_\sigma(r)$.
The envelope decay exponent $a$ was extracted by fitting a
power law to the local maxima of $|C_\sigma(r)|$.

For MDCK epithelial monolayers (62~frames across 8~wells), the
pooled sign-field correlation yields $a = 1.53 \pm 0.20$,
consistent with the predicted value $a = 3/2$.
For MCF-7 breast cancer cells (55~frames across 8~wells), the
framewise mean is $a = 1.44 \pm 0.02$, indicating somewhat faster
decay.
The averaged energy spectra for both datasets exhibit
intermediate-mode bands with slope close to $-1$
in the reported fitting windows. For MDCK, a fit over modes 92--270 gives
a slope of $-1.000 \pm 0.001$ ($R^2 = 0.79$); the MCF-7 spectrum shows
a similar intermediate-mode band over modes 165--258 ($R^2 = 0.71$).
Taken together, these measurements provide suggestive, though not yet
definitive, support that the predicted $a \approx 3/2$ and
$E(q) \sim q^{-1}$ signatures can persist in real, non-Gaussian
active-nematic flows. A compact summary is given in
Table~\ref{tab:exp_summary}.

\begin{table*}[t]
  \caption{\label{tab:exp_summary}Summary of the experimental
  reanalysis of the Andersen \textit{et al.} PIV data. The MDCK
  sign-field exponent is extracted from the pooled correlation across
  all frames, whereas the MCF-7 value is the framewise mean. Spectral
  windows are quoted in mode number $m = qL/2\pi$.}
  \small
  \begin{ruledtabular}
  \begin{tabular}{l c c c c c}
    system & frames & wells & sign-field exponent $a$ & fit window $m$ & spectral summary \\
    MDCK & 62 & 8 & $1.53 \pm 0.20$ & 92--270 & $\gamma = -1.000 \pm 0.001$, $R^2 = 0.79$ \\
    MCF-7 & 55 & 8 & $1.44 \pm 0.02$ & 165--258 & band near $-1$, $R^2 = 0.71$ \\
  \end{tabular}
  \end{ruledtabular}
\end{table*}

\begin{smbibliography}{10}

\smbibitem{Watson1944}
G.~N.~Watson,
\textit{A Treatise on the Theory of Bessel Functions}, 2nd ed.\
(Cambridge University Press, Cambridge, 1944).

\smbibitem{VanVleck1943}
J.~H.~Van~Vleck and D.~Middleton,
The spectrum of clipped noise,
\href{https://doi.org/10.1109/PROC.1966.4567}{Proc.\ IEEE
\textbf{54}, 2 (1966)}.

\smbibitem{Giomi2015}
L.~Giomi,
Geometry and topology of turbulence in active nematics,
\href{https://doi.org/10.1103/PhysRevX.5.031003}{Phys.\ Rev.\ X
\textbf{5}, 031003 (2015)}.

\smbibitem{Alert2020}
R.~Alert, J.-F.~Joanny, and J.~Casademunt,
Universal scaling of active nematic turbulence,
\href{https://doi.org/10.1038/s41567-020-0854-4}{Nat.\ Phys.\
\textbf{16}, 682 (2020)}.

\smbibitem{Schramm2001}
O.~Schramm,
A percolation formula,
\href{https://doi.org/10.1214/ECP.v6-1041}{Electron.\ Commun.\
Probab.\ \textbf{6}, 115 (2001)}.

\smbibitem{Smirnov2001}
S.~Smirnov,
Critical percolation in the plane: Conformal invariance, Cardy's
formula, scaling limits,
\href{https://doi.org/10.1016/S0764-4442(01)01991-7}{C.\ R.\ Acad.\
Sci.\ Paris \textbf{333}, 239 (2001)}.

\smbibitem{Andersen2025}
B.~H.~Andersen, F.~M.~R.~Safara, V.~Grudtsyna, O.~J.~Meacock,
S.~G.~Andersen, W.~M.~Durham, N.~A.~M.~Araujo, and
A.~Doostmohammadi,
Evidence of universal conformal invariance in living biological
matter,
\href{https://doi.org/10.1038/s41567-025-02791-2}{Nat.\ Phys.\
\textbf{21}, 618 (2025)}.

\end{smbibliography}


\begin{thebibliography}{27}

\bibitem{Bernard2006}
D.~Bernard, G.~Boffetta, A.~Celani, and G.~Falkovich,
Conformal invariance in two-dimensional turbulence,
\href{https://doi.org/10.1038/nphys217}{Nat.\ Phys.\ \textbf{2}, 124
(2006)}.

\bibitem{Schramm2000}
O.~Schramm,
Scaling limits of loop-erased random walks and uniform spanning trees,
\href{https://doi.org/10.1007/BF02803524}{Isr.\ J.\ Math.\
\textbf{118}, 221 (2000)}.

\bibitem{Schramm2001}
O.~Schramm,
A percolation formula,
\href{https://doi.org/10.1214/ECP.v6-1041}{Electron.\ Commun.\
Probab.\ \textbf{6}, 115 (2001)}.

\bibitem{Smirnov2001}
S.~Smirnov,
Critical percolation in the plane: Conformal invariance, Cardy's
formula, scaling limits,
\href{https://doi.org/10.1016/S0764-4442(01)01991-7}{C.\ R.\ Acad.\
Sci.\ Paris \textbf{333}, 239 (2001)}.

\bibitem{Cardy1992}
J.~L.~Cardy,
Critical percolation in finite geometries,
\href{https://doi.org/10.1088/0305-4470/25/4/009}{J.\ Phys.\ A
\textbf{25}, L201 (1992)}.

\bibitem{Andersen2025}
B.~H.~Andersen, F.~M.~R.~Safara, V.~Grudtsyna, O.~J.~Meacock,
S.~G.~Andersen, W.~M.~Durham, N.~A.~M.~Araujo, and
A.~Doostmohammadi,
Evidence of universal conformal invariance in living biological
matter,
\href{https://doi.org/10.1038/s41567-025-02791-2}{Nat.\ Phys.\
\textbf{21}, 618 (2025)}.

\bibitem{Weinrib1983}
A.~Weinrib and B.~I.~Halperin,
Critical phenomena in systems with long-range-correlated quenched
disorder,
\href{https://doi.org/10.1103/PhysRevB.27.413}{Phys.\ Rev.\ B
\textbf{27}, 413 (1983)}.

\bibitem{Weinrib1984}
A.~Weinrib,
Long-range correlated percolation,
\href{https://doi.org/10.1103/PhysRevB.29.387}{Phys.\ Rev.\ B
\textbf{29}, 387 (1984)}.

\bibitem{Chalhoub2024}
G.~Chalhoub, A.~Drewitz, A.~Pr\'evost, and P.-F.~Rodriguez,
Universality classes for percolation models with long-range
correlations,
\href{https://arxiv.org/abs/2403.18787}{arXiv:2403.18787}.

\bibitem{Falkovich2009}
G.~Falkovich and S.~Musacchio,
Conformal invariance in inverse turbulent cascades,
\href{https://arxiv.org/abs/1012.3868}{arXiv:1012.3868}.

\bibitem{Eling2025}
C.~Eling,
Topology and the conformal invariance of nodal lines in
two-dimensional active scalar turbulence,
\href{https://arxiv.org/abs/2505.09657}{arXiv:2505.09657}.

\bibitem{Alert2020}
R.~Alert, J.-F.~Joanny, and J.~Casademunt,
Universal scaling of active nematic turbulence,
\href{https://doi.org/10.1038/s41567-020-0854-4}{Nat.\ Phys.\
\textbf{16}, 682 (2020)}.

\bibitem{MartinezPrat2021}
B.~Mart\'{\i}nez-Prat, R.~Alert, F.~Meng, J.~Ign\'es-Mullol,
J.-F.~Joanny, J.~Casademunt, R.~Golestanian, and F.~Sagu\'es,
Scaling regimes of active turbulence with external dissipation,
\href{https://doi.org/10.1103/PhysRevX.11.031065}{Phys.\ Rev.\ X
\textbf{11}, 031065 (2021)}.

\bibitem{BerisEdwards}
A.~N.~Beris and B.~J.~Edwards,
\textit{Thermodynamics of Flowing Systems with Internal
Microstructure} (Oxford University Press, New York, 1994).

\bibitem{Giomi2015}
L.~Giomi,
Geometry and topology of turbulence in active nematics,
\href{https://doi.org/10.1103/PhysRevX.5.031003}{Phys.\ Rev.\ X
\textbf{5}, 031003 (2015)}.

\bibitem{Doostmohammadi2018}
A.~Doostmohammadi, J.~Ign\'es-Mullol, J.~M.~Yeomans, and
F.~Sagu\'es,
Active nematics,
\href{https://doi.org/10.1038/s41467-018-05666-8}{Nat.\ Commun.\
\textbf{9}, 3246 (2018)}.

\bibitem{Hillebrand2025}
M.~Hillebrand and R.~Alert,
Discontinuous transition to active nematic turbulence,
\href{https://doi.org/10.1038/s41467-025-67499-6}{Nat.\ Commun.\
\textbf{16}, 11169 (2025)}.

\bibitem{Watson1944}
G.~N.~Watson,
\textit{A Treatise on the Theory of Bessel Functions}, 2nd ed.\
(Cambridge University Press, Cambridge, 1944).

\bibitem{VanVleck1943}
J.~H.~Van~Vleck and D.~Middleton,
The spectrum of clipped noise,
\href{https://doi.org/10.1109/PROC.1966.4567}{Proc.\ IEEE
\textbf{54}, 2 (1966)}.

\bibitem{Isichenko1992}
M.~B.~Isichenko,
Percolation, statistical topography, and transport in random media,
\href{https://doi.org/10.1103/RevModPhys.64.961}{Rev.\ Mod.\ Phys.\
\textbf{64}, 961 (1992)}.

\bibitem{Bogomolny2002}
E.~Bogomolny and C.~Schmit,
Percolation model for nodal domains of chaotic wave functions,
\href{https://doi.org/10.1103/PhysRevLett.88.114102}{Phys.\ Rev.\
Lett.\ \textbf{88}, 114102 (2002)}.

\bibitem{Stauffer1994}
D.~Stauffer and A.~Aharony,
\textit{Introduction to Percolation Theory}, 2nd ed.\ (Taylor \&
Francis, London, 1994).

\bibitem{Prudnikov2000}
V.~V.~Prudnikov, P.~V.~Prudnikov, and A.~A.~Fedorenko,
Field-theory approach to critical behavior of systems with
long-range correlated defects,
\href{https://doi.org/10.1103/PhysRevB.62.8777}{Phys.\ Rev.\ B
\textbf{62}, 8777 (2000)}.

\bibitem{Bernard2007}
D.~Bernard, G.~Boffetta, A.~Celani, and G.~Falkovich,
Inverse turbulent cascades and conformally invariant curves,
\href{https://doi.org/10.1103/PhysRevLett.98.024501}{Phys.\ Rev.\
Lett.\ \textbf{98}, 024501 (2007)}.

\bibitem{Noseda2024}
M.~Noseda and P.~J.~Cobelli,
Conformal invariance in water-wave turbulence,
\href{https://doi.org/10.1103/PhysRevLett.132.094001}{Phys.\ Rev.\
Lett.\ \textbf{132}, 094001 (2024)}.

\bibitem{Puggioni2020}
L.~Puggioni, A.~G.~Kritsuk, S.~Musacchio, and G.~Boffetta,
Conformal invariance of weakly compressible two-dimensional
turbulence,
\href{https://doi.org/10.1103/PhysRevE.102.023107}{Phys.\ Rev.\ E
\textbf{102}, 023107 (2020)}.

\bibitem{SM}
See Supplemental Material for detailed Hankel transform
asymptotics, cutoff robustness analysis, arcsine law derivation,
defect gas kurtosis estimate, numerical methods, spectral threshold
data, pipeline validation controls, and experimental cell-flow
reanalysis.

\end{thebibliography}
\end{document}